\documentstyle[aps,preprint]{revtex}
\begin{document}
\draft
%
%\tightenlines  --->  for submission to LANL xxx server !!!
%
% >>>>>>>>>>>>>>>>>>>>>>>>>>>>>>>>>>>>>>>>>>>>>>>>>>>>>>>>>>>>>>>>>>>>
% TITLE AND AUTHORS.
%

\title{Surface incompressibility  from
semiclassical \\ Relativistic Mean Field calculations}

\author{S. K. Patra\footnote{Present address:
             Institute of Physics, Sachivalaya Marg,
                Bhubaneswar-{\sl 751 005}, India},
M. Centelles, X. Vi\~nas and M. Del Estal}

\address{Departament d'Estructura i Constituents de la Mat\`eria,
     Facultat de F\'{\i}sica,
\\
     Universitat de Barcelona,
     Diagonal {\sl 647}, E-{\sl 08028} Barcelona, Spain}

\maketitle

% >>>>>>>>>>>>>>>>>>>>>>>>>>>>>>>>>>>>>>>>>>>>>>>>>>>>>>>>>>>>>>>>>>>>
% ABSTRACT, PACS.
%
\begin{abstract}
By using the scaling method and the Thomas-Fermi and Extended
Thomas-Fermi approaches to Relativistic Mean Field Theory the surface
contribution to the leptodermous expansion of the finite nuclei
incompressibility $K_A$ has been self-consistently computed. The
validity of the simplest expansion, which contains volume,
volume-symmetry, surface and Coulomb terms, is examined by comparing
it with self-consistent results of $K_A$ for some currently used
non-linear $\sigma-\omega$ parameter sets. A numerical estimate of
higher-order contributions to the leptodermous expansion, namely the
curvature and surface-symmetry terms, is made.
\end{abstract}

\vspace*{1cm}


%{\it Keywords:}  Relativistic mean field theory, Nuclear 
%incompressibility, Scaling,
%Semiclassical methods, Thomas--Fermi theory

\pagebreak

% >>>>>>>>>>>>>>>>>>>>>>>>>>>>>>>>>>>>>>>>>>>>>>>>>>>>>>>>>>>>>>>>>>>>
% TEXT

The curvature of the nuclear matter equation of state (EOS), i.e.,
the nuclear matter incompressibility $K_{\infty}$ is a key quantity
in nuclear physics because it is related to many properties of
nuclei (such as radii, masses and giant resonances), heavy-ion
collisions, neutron stars and supernova collapses. One
important source of information on $K_{\infty}$ is provided by the
study of the isoscalar giant monopole resonance (GMR) (breathing mode) 
in finite nuclei. In the non-relativistic frame, theoretical
microscopic calculations based on the random-phase approximation
(RPA) \cite{r2} and approximations to it such as the scaling method
\cite{r3,r5,r4} or constrained calculations \cite{r5,r4,r26} using Skyrme
\cite{r5} and Gogny \cite{r6} effective forces lead to a nuclear matter 
incompressibility coefficient $K_{\infty}$ of $215\pm15$ MeV
\cite{r6,r6a}.
A similar analysis carried out within the relativistic
mean-field (RMF) theory with non-linear $\sigma-\omega$ effective
Lagrangians gives a value of $K_{\infty}$ slightly 
higher: $250-270$ MeV \cite{r7}.

The nuclear matter incompressibilty $K_{\infty}$ is not a directly
measurable quantity, what is measured
is, actually, the energy $E_{M}$ of the GMR of finite nuclei. It is 
convenient to write
this energy in terms of the incompressibility $K_A$ for a finite
nucleus of mass number $A$ as:

\begin{equation}
E_{M} = 
\sqrt{\frac{\hbar^2K_{A}}{M<r^2>}},
\label{eqFN1}
\end{equation}
where $<r^2>$ is the rms matter radius and $M$ the nucleon mass. 
The finite nucleus incompressibilty $K_A$ can be parametrized by
means of a leptodermous expansion \cite{r3} which is similar to the
liquid drop mass formula:

\begin{equation}
K_{A} = K_{\infty}+K_{sf}A^{-1/3}+K_{vs}I^2+K_{coul}Z^2A^{-4/3}+.....,
\label{eqFN2}\end{equation}
where $I=(N-Z)/A$ is the neutron excess. Eq. (\ref{eqFN2}) suggests that
it is possible to fit the coefficients of the expansion to the
experimental data in a
model independent way. Although some effort along these lines has been
made in the past \cite{r8}, the fact that a fit of the parameters
of  Eq. (\ref{eqFN2}) to experimental data does not lead to a unique 
determination of the parameters is well established 
\cite{r6,r9,r10}. Rather, the nuclear matter 
incompressibility has to be determined from effective forces which 
reproduce, in a microscopic calculation, the experimental values of the
GMR excitation energy in heavy nuclei \cite{r6}. 

It is also possible to fit $K_A$ calculated 
microscopically within the scaling model for a given effective
interaction to the leptodermous expansion
Eq.  (\ref{eqFN2}). This has been done, for example, in the non-relativistic
frame using Skyrme forces \cite{r11}. In this case the coefficients
entering Eq.  (\ref{eqFN2}) can be expressed through infinite
and semi-infinite nuclear matter properties calculated
with the Hartree-Fock approximation for each considered interaction.
In particular, the volume-symmetry ($K_{vs}$) and Coulomb
($K_{coul}$) coefficients depend on some parameters of the 
liquid droplet model \cite{r15} computed only using nuclear matter
properties \cite{r3}. The surface coefficient $K_{sf}$, also derived
in \cite{r3}, can be  written as \cite{r19}:
\begin{equation}
K_{sf} = 4\pi r_0^2\left [ \left( 22 + \frac{54}{K_{\infty}}\rho_0^3
{\stackrel{...}{e}}_{\infty}(\rho_0) \right) \sigma(\rho_0) 
+ 9\rho_0^2{\ddot\sigma} (\rho_0) \right ].
\label{eqFN3}\end{equation}
The surface tension $\sigma$ is calculated in
symmetric semi-infinite nuclear matter and is defined as:
\begin{equation}
\sigma(\rho_c) = \int_{-\infty}^{+\infty} \left\{ {\cal H}(\rho)
-e_{\infty}(\rho_c) \rho \right\} dz, 
\label{eqFN4}\end{equation}
where $\rho$ is the density profile whose central value is given
by $\rho_c=\rho({-\infty})$, ${\cal H}$ is the energy density and
$e_{\infty}$ is the energy per particle in nuclear matter at density $\rho_c$.
In Eq. (\ref{eqFN3}) dots indicate the
derivatives with respect to the central density and all
the quantities are evaluated at a central density equal to the
nuclear matter saturation density $\rho_0$,
which is related to the radius constant $r_0$ through
$4\pi r_0^3\rho_0/3=1$.

The key quantity entering Eq. (\ref{eqFN3}) is $\ddot\sigma$ which
is the second derivative of $\sigma(\rho_c)$ with respect to $\rho_c$
calculated 
at $\rho_c=\rho_0$. The determination of $\ddot\sigma$ also requires
knowledge of how the density profile $\rho$ is modified during 
compression \cite{r20}. In the study of the breathing mode a scaling 
transformation of the densities is assumed. Actually, the coefficients
entering the parametrization
(\ref{eqFN2}) can be derived under this hypothesis \cite{r3}. 
The scaling transformation
means that the density changes according to the transformation
$\mbox{\boldmath $r$}\rightarrow\lambda \mbox{\boldmath $r$}$ 
and consequently

\begin{eqnarray}
\rho_\lambda(\mbox{\boldmath$r$}) & = &
\lambda^3 \rho(\lambda\mbox{\boldmath$r$}).
\label{eqFN5}\end{eqnarray}
Thus, in the scaling approach:

\begin{equation}
\ddot\sigma(\rho_0)=\left[\frac{d^2\sigma(\rho_c)}{d\rho_c^2}\right]_{\rho_0}
= \frac{1}{9\rho_0^2}\left[\frac{d^2\sigma}{d\lambda^2}\right]_{\lambda=1}.
\label{eqFN6}\end{equation}
To obtain the surface incompressibility coefficient $K_{sf}$ for a 
given effective interaction, it is necessary, first of all, to calculate
the scaled surface tension $\sigma_{\lambda}$ by replacing the densities by 
the scaled densities given by Eq. (\ref{eqFN5}) in Eq. (\ref{eqFN4}). In the
non-relativistic frame this can be easily done within the Hartree-Fock 
scheme using zero-range Skyrme forces and a simple analytical expression
for $\sigma_{\lambda}$ is obtained \cite{r11,r19}.

The self-consistent calculation of $K_{sf}$ within the RMF
approximation using the $\sigma-\omega$ model is more involved due to
the problem of the change in the meson fields induced by the scaled
nuclear densities \cite{r12}. To our knowledge, only approximate
calculations of $K_{sf}$ have been developed in the past for the
relativistic model. This is the
case of the Relativistic Thomas-Fermi (RTF) calculations of Refs.
\cite{r12,r24b} where a local density approximation of the meson
fields was used. Another approach is related with the study of nuclei
under an external pressure. Starting from a schematic energy density
functional and adding a density-dependent constraint which simulates
the pressure, analytical expressions for the surface tension $\sigma$
as a function of the bulk density $\rho_c$ can be derived for a wide
class of compression modes, in particular, for the scaling mode
\cite{r20}. This way one obtains the following formula for
$\ddot{\sigma}$ in the scaling mode:
\begin{equation}
\ddot\sigma(\rho_0)=-\frac{19}{81} \frac{K_{\infty}\alpha}{\rho_0}, 
\label{eqFN7}
\end{equation}
where $\alpha$ is the surface diffuseness parameter of a symmetric
Fermi density. This pocket formula has been employed to
estimate $K_{sf}$ in the RMF model for several non-linear
$\sigma-\omega$ parameter sets \cite{r14}. A symmetric Fermi function
that reproduces in the best way the density profile obtained from a
Hartree calculation of semi-infinite nuclear matter has been used in
ref. \cite{r14} to determine the $\alpha$ parameter of Eq.
(\ref{eqFN7}). 

Very recently, the scaling method applied to the RMF theory in the RTF
and Relativistic Extended Thomas-Fermi (RETF) approaches has been used
to self-consistently obtain the excitation energy of the GMR of finite
nuclei \cite{patra01,r21}. Our aim in the present paper is, firstly to
obtain the surface coefficient $K_{sf}$ self-consistently in the RTF
and RETF approaches developed in \cite{patra01,r21} for some linear
and non-linear $\sigma-\omega$ parameter sets. On the other hand, we
want to check whether the leptodermous expansion of the finite nucleus
incompressibility Eq. (\ref{eqFN2}) can reproduce the corresponding
fully self-consistent value obtained in the RETF approach \cite{r21}
with some selected non-linear $\sigma-\omega$ parameter sets.

 The key point of our 
semiclassical approach is that the local Fermi momentum 
$k_F$ and the effective mass $m^*$ scale as \cite{patra01,r21}:
\begin{equation}
{k_F}_\lambda =
\lambda k_F(\lambda\mbox{\boldmath $r$}), \qquad m^*_\lambda(\mbox{\boldmath$r$}) = 
 \lambda {\tilde m}^* (\lambda\mbox{\boldmath$r$}),  
\label{eqFN8}
\end{equation}
where $\tilde m^*$ is still a function of $\lambda$.
With the help of Eq. (\ref{eqFN8}), the nuclear part of the energy and
the scalar density including $\hbar^2$ corrections, which are functionals
of $k_F$ and $m^*$, scale as:
\begin{eqnarray}
{\cal E}_\lambda(\mbox{\boldmath$r$}) & = &
\lambda^4 {\tilde{\cal E}} (\lambda\mbox{\boldmath$r$}), \qquad
\rho_{s\lambda} (\mbox{\boldmath$r$}) =
\lambda^3 {\tilde\rho}_{s} (\lambda\mbox{\boldmath$r$}).
\label{eqFN9}
\end{eqnarray}
Again $\tilde{\cal E}$ and $\tilde\rho_s$
are functions of the collective coordinate $\lambda$ because of
their dependence on  ${\tilde m}^*$.
Thus the scaled surface tension can be written as
\cite{patra01,r21,r22}:
\begin{eqnarray}
\sigma_{\lambda} & = &
\int \left [ \lambda^4 {\cal H}_{\lambda}(\lambda z)
-e_{\infty}(\lambda^3\rho_0)\lambda^3\rho(\lambda z)\right ] 
\frac{d(\lambda z)}{\lambda} 
\nonumber \\[3mm]
 & & \mbox{}
=\int d(\lambda z)\left \{\lambda^3\tilde{\cal E}
+\lambda^2g_{v}V_{\lambda}\rho-\frac{1}{2}\lambda
\left [ (\mbox{\boldmath$\nabla$} V_{\lambda})^2+\frac{m_{v}^2}{\lambda^2}V_{\lambda}^2\right ]
+\frac{1}{2}\lambda
\left [ (\mbox{\boldmath$\nabla$} \phi_{\lambda})^2
+\frac{m_{s}^2}{\lambda^2}\phi_{\lambda}^2\right ]\right .
\nonumber \\[3mm]
 & & \mbox{}
\left .
+\frac{b\phi_{\lambda}^3}{3\lambda}
+\frac{c\phi_{\lambda}^4}{4\lambda}
-\lambda^2e_{\infty}(\lambda^3\rho_0) \rho\right \},
\label{eqFN10}
\end{eqnarray}
where all densities and fields depend on the variable $\lambda z$.
With the help of the Klein-Gordon equations for the scaled vector
and scalar fields derived from (\ref{eqFN10}), the scaled surface 
tension can be recast as:

\begin{eqnarray}
\sigma_{\lambda} & = &
\int d(\lambda z) \left \{ \lambda^3 \tilde{\cal E}_{\lambda}
+\frac{1}{2}\lambda^2g_{v}V_{\lambda}\rho
+\frac{1}{2}\lambda^2g_{s}\phi_{\lambda}\tilde\rho_{s}
-\frac{b\phi_{\lambda}^3}{6\lambda}
-\frac{c\phi_{\lambda}^4}{4\lambda}
-\lambda^2 e_{\infty}(\lambda^3\rho_0) \rho\right \}.
\label{eqFN11}
\end{eqnarray}
Using the explicit RTF or RETF expressions for
the nuclear part of the energy and for the scalar density
\cite{patra01,r21,r22,r23} together with the
Klein-Gordon equations for $V_{\lambda}$, $\phi_{\lambda}$,
$\frac{\partial V_{\lambda}}{\partial \lambda}$ and
$\frac{\partial \phi_{\lambda}}{\partial \lambda}$ derived from 
Eq. (\ref{eqFN10}), after some algebra the first and second
derivatives of the scaled surface tension $\sigma_{\lambda}$ with respect
to $\lambda$ at $\lambda=1$ read (see Refs. \cite{patra01,r21} for
more details):
\begin{eqnarray}
\left. \frac{d\sigma_{\lambda}}{d\lambda}\right|_{\lambda=1} & =&
2\sigma
+ \int_{-\infty}^{+\infty} dz
\left \{
{\cal E}-\rho_{s} m^*-m_{s}^2\phi^2-\frac{1}{2}g_{s}\rho_{s}\phi
\right .
\nonumber \\[3mm]
 & & \mbox{}
\left .
-\frac{1}{2}b\phi^3-\frac{1}{4}c\phi^4
+\frac{1}{2}g_{v}\rho V+m_{v}^2 V^2 
\right \}
=0
\label{eqFN12}
\end{eqnarray}
and
\begin{eqnarray}
\left.
\frac{d^2\sigma_{\lambda}}{d\lambda^2}\right|_{\lambda=1}& = &
-6\sigma
+ \int_{-\infty}^{+\infty}dz\left\{
b\phi^3-(b\phi^2+2m_{s}^2\phi) \frac{\partial \phi_{\lambda}}
{\partial \lambda}|_{\lambda=1}+3m_{s}^2\phi^2
\right .
\nonumber \\[3mm]
& & \mbox{}
\left .
+2m_{v}^2V\frac{\partial V_{\lambda}}{\partial
\lambda}|_{\lambda=1}
-3m_{v}^2V^2+m\frac{\delta\rho_{s}}{\delta m^*}
\left(m^*+g_{s}\frac{\partial \phi_{\lambda}}{\partial\lambda}|_{\lambda=1}
\right)
-K_{\infty}\rho\right \}.
\label{eqFN13}
\end{eqnarray}
The first derivative of $e_{\infty}(\lambda^3\rho_0)$ at
$\lambda=1$ is just three times the pressure calculated at saturation
density and thus it vanishes, while the second derivative 
gives $K_{\infty}\rho$ \cite{r21,r24}. On the other hand,
since in the self-consistent RTF and RETF calculations the
inputs for computing Eqs. (\ref{eqFN12})-(\ref{eqFN13}) are
quantities obtained from the solution of the variational
equations associated with the surface tension (\ref{eqFN10})
at $\lambda=1$, the so-called "sigma dot" theorem is rigorously fulfilled
\cite{r24a}. The method therefore allows $\ddot{\sigma}$ and consequently 
$K_{sf}$ to be computed on top of a self-consistent RTF or RETF
calculation of the surface tension in symmetric semi-infinite
nuclear matter. This is similar to what happens in the non-relativistic
frame with Skyrme forces \cite{r19}, although in the relativistic case 
additional Klein-Gordon equations for 
$\frac{\partial V_{\lambda}}{\partial\lambda}$ and 
$\frac{\partial\phi_{\lambda}}{\partial\lambda}$ at $\lambda=1$
have to be solved.

Now we shall discuss the results obtained from the self-consistent RTF
and RETF methods in the scaling approximation. Table 1 collects
$K_{\infty}$, $\ddot{\sigma}$ and $K_{sf}$ for the non-linear NL-Z2
\cite{rx1}, NL1 \cite{r16},NL3 \cite{r17}, NL-RA1 \cite{rx2}, NL-SH
\cite{r18} and NL2 \cite{rx3} and the linear HS \cite{rx4} and L1
\cite{rx3} parameter sets. One observes that in both the RTF and RETF
calculations $\ddot{\sigma}$ and $K_{sf}$ decrease (become more
negative) with increasing bulk incompressibility $K_{\infty}$. The RTF
and RETF values of $\ddot{\sigma}$ and $K_{sf}$ for a given parameter
set are, in general, rather different from one another, which means
that the precise value of these quantities is model dependent. This is
known to happen also with other quantities related with the nuclear
surface. For example, such is the case of the surface energy
coefficient of the leptodermous expansion of the binding energy of a
nucleus, which is calculated as $4\pi r_0^2 \sigma$. The quality of
the RTF and RETF approximations for semi-infinite nuclear matter and
finite nuclei with respect to the RMF Hartree approach, and its
dependence on the effective interaction, was investigated in Refs.
\cite{r22,centelles92,speicher93} by analyzing the results obtained
with many different parameter sets. It was noticed that the RTF
results fluctuate around the corresponding Hartree results. Due to
this fact there exist parametrizations for which the RTF approximation
agrees by chance with the Hartree result. The behaviour of RETF
results in comparison with the Hartree solutions was found to be less
dependent on the parameters of the force than in the RTF case, and it
turned out that on the average the RETF results are in better
agreement with the Hartree ones \cite{r22,centelles92,speicher93}.

The first contribution to $K_{sf}$ in Eq. (\ref{eqFN3}) comes from the
surface tension, let us call it $K_{sf}^\sigma$. The deviation found
in the value of the surface tension from RTF calculations with respect
to the corresponding RMF Hartree calculations is strongly correlated
with the value of the effective mass in nuclear matter $m^*_\infty/m$
\cite{r22,speicher93}. For small values of $m^*_\infty/m$ the RTF
surface tension is larger than the Hartree one, it practically agrees
with the Hartree result for $m^*_\infty/m \sim 0.65$, and it becomes
smaller than the Hartree result for larger $m^*_\infty/m$. On the
other hand, the RETF result for the surface tension is consistently
lower than the Hartree result and much less dependent on the specific
value of $m^*_\infty/m$. (A similar situation is found for the total
energy of finite nuclei \cite{r22,centelles92,speicher93}.) These
trends, of course, are also reflected in $K_{sf}^\sigma$. For example,
for NL1 ($m^*_\infty/m= 0.57$) we have $K_{sf}^\sigma= 402.6$, 377 and
429.3 MeV in the Hartree, RETF and RTF approaches, respectively. For
NL2 ($m^*_\infty/m= 0.67$) it is $K_{sf}^\sigma= 479.6$, 439.1 and
465.7 MeV in the Hartree, RETF and RTF calculation, respectively.

The second contribution to $K_{sf}$ in Eq. (\ref{eqFN3}) is due to the
second derivative of the surface tension. The results for
$\ddot{\sigma}$ in the RTF approach decrease with $K_\infty$ faster
than in the RETF calculation. At small values of $K_\infty$ the RTF
value of $\ddot{\sigma}$ is less negative than that computed in the
RETF approach, while the opposite happens for higher values of
$K_\infty$. Both approaches predict the same value of $\ddot{\sigma}$
for an incompressibility around that of NL1 (211 MeV). A similar
behaviour is displayed by the self-consistent values of $K_{sf}$,
although the crossing point between the RTF and RETF predictions is
shifted to a larger value of $K_\infty$ (around 280 MeV) due to the
fact that the contribution proportional to $\sigma$ ($K_{sf}^\sigma$)
is larger in the RTF approach than in the RETF approach for the
parameter sets considered here. 

The $\ddot{\sigma}$ values obtained from the pocket formula Eq.
(\ref{eqFN7}) using the surface diffuseness of the RTF or RETF
semi-infinite nuclear matter density profiles also decrease with
$K_{\infty}$, though the estimate provided by Eq. (\ref{eqFN7}) does
not reproduce in general the self-consistent values very precisely.
The approximate $\ddot{\sigma}$ is always smaller when calculated from
the RTF approach than from the RETF approach for the parametrizations
of Table 1. Using Eq. (\ref{eqFN7}) to estimate the value of $K_{sf}$
in first approximation, one finds that the RETF result is closer to
the Hartree value than the RTF result for the sets NL-Z2, NL1, NL3 and
NL-RA1. For NL2 and L1 it is the RTF estimate which lies closer to the
Hartree estimate. For NL-SH and HS the approximate Hartree prediction
lies roughly in between of the approximate RTF and RETF values. To the
extent that Eq. (\ref{eqFN7}) is applicable, it provides a hint of
where the unknown exact Hartree value for $K_{sf}$ should lie with
respect to the self-consistent RTF and RETF results presented in Table
1. 

Another different approach to computing $K_{sf}$ was proposed in Refs.
\cite{r12,r24b}. It is based on the scaling method together with a
local density approximation for the meson fields within the RTF
approach. In Ref. \cite{r24b} a $K_{sf}$ of approximately $-1000$ MeV
was reported for a linear set with $K_{\infty}=545$ MeV (similar to
the HS set). This result is in good agreement with that of our
self-consistent RTF calculation for the HS set, though it is clearly
larger in absolute value than the RETF result for $K_{sf}$ obtained
with the HS set. On the other hand, our self-consistent RTF and RETF
surface incompressibilities differ considerably from the estimate of
Ref. \cite{r25} where approximate $K_{sf}$ values of $-333.1$ and
$-610.1$ MeV were reported for the NL1 and NL-SH parametrizations
calculated with the method used in \cite{r12}.

It should also be pointed out that in our self-consistent
semiclassical calculations we find that the ratio between the surface
and bulk incompressibilities increases with $K_{\infty}$ (in agreement
with the results of \cite{r24b}). In the RETF case this ratio is close
to one, as happens for the non-relativistic Skyrme forces \cite{r26},
provided that the bulk incompressibility $K_{\infty}$ of the
interaction is not excessively high. In the RTF case the ratio between
the surface and bulk incompressibilities increases much faster with
$K_{\infty}$ than in the RETF calculations, and it considerably
differs from unity for parametrizations with either a very low or a
very high bulk incompressibility. In Figure 1 we plot $-K_{sf}$ as a
function of $K_{\infty}$ for the parameter sets considered in Table 1.
As in the non-relativistic case \cite{r3}, $K_{sf}$ varies roughly
linearly with $K_{\infty}$. A linear fit of all the points gives
$-K_{sf} = 1.47 K_{\infty} - 84$ in the RETF model and $-K_{sf} = 2.19
K_{\infty} - 295$ in the RTF model. If only the non-linear
parametrizations are included in the fit one obtains $-K_{sf} = 1.35
K_{\infty} - 54$ and $-K_{sf} = 1.96 K_{\infty} - 238$ in the RETF and
RTF cases, respectively.

The surface incompressibility coefficient is both large and negative,
thus its contribution considerably reduces the finite nucleus
incompressibility $K_A$ with respect to the nuclear matter limit
$K_{\infty}$. This result, although obtained in the scaling model,
illustrates  the physical effect that the compression of the surface
provides a considerable reduction of $K_A$, which is also found in
more fundamental RPA calculations \cite{r6}. In Ref. \cite{r21} we
have  self-consistently computed the finite nucleus incompressibility
$K_A$  using the RETF approach and the scaling method which we have
employed  in the present work to obtain $K_{sf}$. Thus we can now
precisely  check the ability of the leptodermous expansion Eq.
(\ref{eqFN2}) in reproducing the full calculation of $K_A$ carried out
in Ref. \cite{r21} for various finite nuclei.

The coefficients $K_{vs}$ and $K_{coul}$ entering Eq. (\ref{eqFN2})
are computed using nuclear matter properties only. Explicit
expressions for these coefficients in the non-linear $\sigma-\omega$
model are reported in \cite{r13}. In our analysis we will use the NL1,
NL3 and NL-SH parameter sets for which the numerical values of these
coefficients are given in \cite{r14}. The surface incompressibility
coefficient is the self-consistent value taken from Table 1. Table 2
collects $K_A$ obtained from the full self-consistent RETF calculation
\cite{r21} as well as the value $K(A,I)$ given by Eq. (\ref{eqFN2})
for $^{40}$Ca, $^{48}$Ca, $^{56}$Ni, $^{90}$Zr, $^{116}$Sn,
$^{132}$Sn, $^{144}$Sm and $^{208}$Pb. From this Table it can be seen
that the leptodermous expansion with the terms given in Eq.
(\ref{eqFN2}) fails to describe small nuclei and also very asymmetric
nuclei such as $^{132}$Sn or $^{208}$Pb. In addition, some words of
caution should be said about the Coulomb term in Eq. (\ref{eqFN2}). In
the self-consistent scaling calculation of the finite nucleus
incompressibility, the Coulomb energy does not participate directly if
the  scaling Eq. (\ref{eqFN5}) for the density is assumed to apply
\cite{r5,r21}.  Thus, the Coulomb term in Eq. (\ref{eqFN2}) should be
related to the change in $K_{A}$ when the Coulomb interaction is
switched off in the self-consistent calculation. The Coulomb term in
Eq. (\ref{eqFN2}) overestimates this change by approximately 6 MeV for
NL1, 3 MeV for NL3 and 1 MeV for NL-SH.

Now we would like to analyze whether the addition of some higher order
terms in the
leptodermous expansion  Eq. (\ref{eqFN2}) improves the agreement with the
$K_A$ results calculated self-consistently. 
In particular, we will focus our attention on the
curvature $K_{cv}A^{-1/3}$ and the surface-symmetry $K_{ss}I^2A^{-1/3}$
terms. Although these terms could be derived by enlarging the
leptodermous expansion of Blaizot \cite{r3}, as has been done in the
non-relativistic case \cite{r11}, it becomes more complicated in the 
relativistic case. Thus, for a fast estimate of the curvature and
surface-symmetry terms, we perform a numerical fit. To do this, we
follow the same strategy as in Ref. \cite{r11}. First
 we consider symmetric nuclei with the Coulomb force switched off. In
this case the leptodermous expansion Eq.(\ref{eqFN2}) 
(adding the curvature term) reduces to:

\begin{eqnarray}
K_A  = K_{\infty}+K_{sf}A^{-1/3}+K_{cv}A^{-2/3}.
\label{eqFN14}
\end{eqnarray}
In Figure 2 we plot $[K_{A}-K_{\infty}]A^{1/3}$ versus $A^{-1/3}$ for
the three parameters sets used in this analysis. Here
$K_{\infty}$ is the nuclear matter incompressibility given in
Table 1 and $K_{A}$ are the
self-consistent incompressibilities calculated
 for $A$ ranging from 300 up to 300000. In the linear fit of these curves
the $y-$axis intercept gives $K_{sf}$ of the
corresponding force, while the slope gives $K_{cv}$. The surface
terms obtained in this way are $-$246.1, $-$328.4 and $-$435.8
MeV for the NL1, NL3 and NL-SH parameter sets, which are very close to
the corresponding self-consistent values  (see Table 1). The 
estimates of the curvature term in the leptodermous expansion of the
finite nucleus incompressibility obtained with NL1, NL3 and NL-SH are
$-$317.2, $-$229.8 and $-$185.6 MeV respectively. 

To obtain the surface-symmetry contribution, we have found it
convenient to parametrize the difference between the self-consistent
incompressibilities $K_A$ of a given nucleus with neutron excess $I$
and the corresponding symmetric nucleus as:
\begin{eqnarray}
K_{A,I} - K_{A,I=0} = K_{vs}I^2+K_{ss}I^2A^{-1/3},
\label{eqFN15}
\end{eqnarray}
where again uncharged nuclei have been considered. For each parameter
set and according to (\ref{eqFN15}), if $[K_{A,I}-K_{A,I=0}]I^{-2}$ is
plotted versus $A^{-1/3}$ a unique curve should be found which is
independent of the value of I. However, one obtains a family of almost
parallel lines whose slope is $K_{ss}$. The splitting of these lines
gives us information on the higher order symmetry contributions missed
in the parametrization (\ref{eqFN15}). Thus we will estimate the
surface-symmetry term from a linear fit of the curve corresponding to
$I=0.1$, which roughly corresponds to an average asymmetry along the
periodic table. This curve is plotted in Figure 3 for $A$ ranging from
250 to 200000 for each considered parameter set. The corresponding
y-axis intercepts agree very well with the $K_{vs}$ values calculated
in nuclear matter ($-$676.1, $-$698.9 and $-$794.5 MeV for NL1, NL3
and NL-SH respectively \cite{r14}). Our estimate of the
surface-symmetry contribution to $K_A$ corresponds to the slopes of
these linear fits, which are 1951.4, 1754.0 and 1716.5 MeV for NL1,
NL3 and NL-SH respectively.

Table 3 collects the self-consistent finite nuclei incompressibility
$K_{A}$ (without Coulomb) compared with the macroscopic
parametrizations $K(A,I)$ (Eq. (\ref{eqFN2})) and $K^*(A,I)$ which
contains the curvature and surface-symmetry contributions obtained
from the previously discussed fits. Again, the self-consistent
incompressibilities corresponding to the lightest nuclei and the very
asymmetric nuclei are not well reproduced by the simplest expansion
Eq. (\ref{eqFN2}). If the curvature and surface-symmetry corrections
are included, the improved macroscopic formula $K^*(A,I)$ reproduces
the self-consistent incompressibilities with an error, on average,
smaller than $1.2\%$, $0.9\%$ and $0.3\%$ for the NL1, NL3 and NL-SH
parameter sets. In order to gain some insight into the accuracy of our
estimate of the curvature and surface-symmetry contributions, we fit the
self-consistent results for the finite nuclei considered in Table 3 to
a leptodermous expansion including curvature and surface-symmetry terms.
The volume, surface and volume symmetry coefficients are taken from
self-consistent infinite and semi-infinite nuclear matter
calculations. The results of this calculation show that the difference
of the curvature contribution obtained from the fit in the asymptotic
region and from finite nuclei is always less than $10\%$, whereas the
difference in the surface-symmetry contribution lies below $3\%$.

We have applied the scaling method in the Thomas-Fermi and Extended
Thomas-Fermi approximations to the Relativistic Mean Field Theory to
self-consistently calculate the surface coefficient $K_{sf}$ of the
leptodermous expansion of the finite nucleus incompressibility derived
within the Blaizot model. The ratio between the surface and bulk
incompressibilities obtained in our semiclassical calculation
increases with the nuclear matter incompressibility, more strongly in
the RTF than in the RETF case. In the RETF calculations this ratio is
close to one, as in the case of non-relativistic Skyrme forces, for
the non-linear parameter sets which have a nuclear matter
incompressibility not larger than roughly 300 MeV.

For the analyzed $\sigma-\omega$ parameter sets, the leptodermous
expansion Eq. (\ref{eqFN2}) is not able to reproduce very precisely
the finite nuclei incompressibilities obtained self-consistently. In
particular the macroscopic contribution of the Coulomb force can
differ from the self-consistent contribution up to 6 MeV. We have
numerically estimated higher order contributions to the leptodermous
expansion, namely curvature and surface-symmetry terms, in the
asymptotic region (i.e., for very large uncharged systems). We have
found that the finite nuclei incompressibilities are reasonably well
reproduced by an extended leptodermous expansion which includes
curvature and surface-symmetry contributions.

\acknowledgments

Useful discussions with E. Vives, M. Farine and J.N. De
are acknowledged.
The authors would like to acknowledge support from the DGICYT (Spain)
under grant PB98-1247 and from DGR (Catalonia) under grant
2000SGR-00024.  S.K.P. thanks the Spanish Education Ministry grant
SB97-OL174874 for financial support and the Departament d'Estructura i
Constituents de la Mat\`eria of the University of Barcelona for kind
hospitality.

% REFERENCES.

%
\pagebreak
%
%
% >>>>>>>>>>>>>>>>>>>>>>>>>>>>>>>>>>>>>>>>>>>>>>>>>>>>>>>>>>>>>>>>>>>>
%
% FIGURE CAPTIONS.
%
\section*{Figure captions}
\begin{description}
\item[Figure 1.]
Surface incompressibility coefficient
versus the nuclear matter incompressibility modulus 
for the parameter sets of Table 1.
\item[Figure 2.]
$(K_A-K_{\infty})A^{1/3}$ versus $A^{-1/3}$ computed for several
uncharged and symmetric nuclei from $A=250$ to $A=300000$ for the
NL1, NL3 and NL-SH parameter sets.
\item[Figure 3.]
$(K_{A,I}-K_{A,I=0})/I^2$  versus $A^{-1/3}$ for several
uncharged nuclei from $A=250$ to $A=200000$ with a neutron excess
 $0.10$ for the NL1, NL3 and NL-SH parameter sets. 
\end{description}
\pagebreak
% >>>>>>>>>>>>>>>>>>>>>>>>>>>>>>>>>>>>>>>>>>>>>>>>>>>>>>>>>>>>>>>>>>>>
% THE TABLES.
%
%\section*{Table 1}
\vspace{2cm}
%\begin{center}  \small
\begin{table}
\caption{Values of $\ddot\sigma$ (in MeV\,fm$^4$) and $K_{sf}$ (in
MeV) calculated with the RTF and RETF approaches and the scaling
method for several parameter sets. The nuclear matter
incompressibility modulus $K_{\infty}$ (in MeV) and the
$-K_{sf}/K_{\infty}$ ratio are also listed.}
\vspace{0.5cm}
\begin{tabular}{cccccccc}
 & & \multicolumn{3}{c}{RTF} & \multicolumn{3}{c}{RETF} \\
     \cline{3-5}               \cline{6-8} 
  & $K_{\infty}$
  & $\ddot\sigma$ & $K_{sf}$ &$-K_{sf}/K_{\infty}$ 
  & $\ddot\sigma$ & $K_{sf}$ &$-K_{sf}/K_{\infty}$ \\
\hline
NL-Z2  & 172.2& $-$113.9 &  $-$85.2 &0.49 &$-$131.2 &$-$182.5 &1.06 \\
NL1    & 211.1& $-$170.3 & $-$170.6 &0.81 &$-$171.8 &$-$225.4 &1.07 \\
NL3    & 271.5& $-$224.2 & $-$310.4 &1.14 &$-$209.3 &$-$313.7 &1.16 \\
NL-RA1 & 285.3& $-$235.5 & $-$335.4 &1.18 &$-$216.6 &$-$326.7 &1.15 \\
NL-SH  & 355.0& $-$292.7 & $-$469.8 &1.32 &$-$258.2 &$-$429.6 &1.21 \\
NL2    & 399.2& $-$295.9 & $-$521.0 &1.31 &$-$279.0 &$-$482.8 &1.21 \\
HS     & 546.8& $-$521.5 & $-$996.7 &1.82 &$-$424.9 &$-$804.2 &1.47 \\
L1     & 625.6& $-$422.6 &$-$1024.6 &1.64 &$-$320.6 &$-$787.1 &1.26 \\
\end{tabular}
\end{table}

%\section*{Table 2}
\vspace{2cm}
%\begin{center}  \small
\begin{table}
\caption{ Finite nuclei incompressibilities (in MeV)
 calculated with the self-consistent RETF
approach $(K_A$) and with the leptodermous expansion Eq. (\ref{eqFN2})
($K(A,I)$). Results are presented for the NL1, NL3 and NL-SH parameter
sets.  }
\vspace{0.5cm}
\begin{tabular}{cccccccc}
  & NL1  & &NL3 & & NL-SH & \\
\hline
  &$K_{A}$& $K(A,I)$ &$K_{A}$ &$K(A,I)$ &$K_{A}$ &$K(A,I)$ & \\
\hline
$^{40}$Ca& 108.2&128.1 & 145.3  & 161.0& 196.8 &208.6 &\\
$^{48}$Ca& 111.1&116.9 & 147.4  & 151.0& 198.3 &198.4 &\\
$^{56}$Ni& 115.0&130.8 & 153.2  & 166.0& 207.1 &216.7 &\\
$^{90}$Zr& 122.5&129.3 & 161.6  & 167.3& 217.5 &221.1 &\\
$^{116}$Sn& 124.3&126.3 & 163.4  & 165.4& 219.8 &220.4 &\\
$^{132}$Sn& 121.3&105.4 & 157.6  & 144.9& 210.9 &197.5 &\\
$^{144}$Sm& 125.4&125.3 & 164.5  & 165.3& 221.6 &221.5 &\\
$^{208}$Pb& 124.1&111.1 & 161.1  & 152.1& 216.7 &208.1 &\\
\end{tabular}
\end{table}
%\end{center}
%
\pagebreak

% >>>>>>>>>>>>>>>>>>>>>>>>>>>>>>>>>>>>>>>>>>>>>>>>>>>>>>>>>>>>>>>>>>>>
%
%\section*{Table 3}
\vspace{2cm}
%\begin{center}  \small
\begin{table}
\caption{ Finite nuclei incompressibilities (in MeV)
for several {\it uncharged}
nuclei calculated self-consistently using the
RETF approach $(K_A)$, with the leptodermous expansion Eq. (\ref{eqFN2})
$(K(A,I))$ and including the curvature and surface-symmetry contributions
$(K^*(A,I))$. Results are presented for the NL1, NL3 and NL-SH parameter
sets.
}
\vspace{0.5cm}
\begin{tabular}{ccccccccccc}
%  & NL1  & &NL2 & & NL3 && NL-SH && \\
  &  & NL1& & & NL3 && &NL-SH& \\
\hline
  & $K_{A}$& $K(A,I)$ &$K^*(A,I)$&$K_{A}$ &$K(A,I)$&$K^*(A,I)$ 
&$K_{A}$ &$K(A,I)$ &$K^*(A,I)$ & \\
\hline
$^{40}$Ca &118.6 &145.2 & 118.1 & 160.1& 179.8 &160.2 & 213.4 & 229.4
& 213.5&\\
$^{48}$Ca &119.6 &130.3 & 121.2 & 159.7& 165.8 &161.8 & 215.1 & 214.7
& 213.8&\\
$^{56}$Ni &129.3 &152.2 & 130.5 & 172.7& 189.6 &173.9 & 230.5 & 242.7
& 230.0&\\
$^{90}$Zr &139.6 &152.5 & 142.0 & 184.2& 192.9 &186.3 & 244.3 & 249.3
& 244.8&\\
$^{116}$Sn &144.0 &152.0 & 146.3 & 189.0& 193.9 &191.1 & 250.1 & 251.8
& 250.7&\\
$^{132}$Sn &137.2 &127.1 & 137.4 & 179.3& 168.9 &180.2 & 236.8 & 223.9
& 236.58&\\
$^{144}$Sm &148.5 &155.0 & 150.7 & 194.3& 198.2 &196.3 & 256.7 & 257.7
& 257.2&\\
$^{208}$Pb &148.4 &142.8 & 148.5 & 193.4& 187.3 &194.0 & 255.0 & 246.9
& 254.6&\\
\end{tabular}
\end{table}
%\end{center}
%
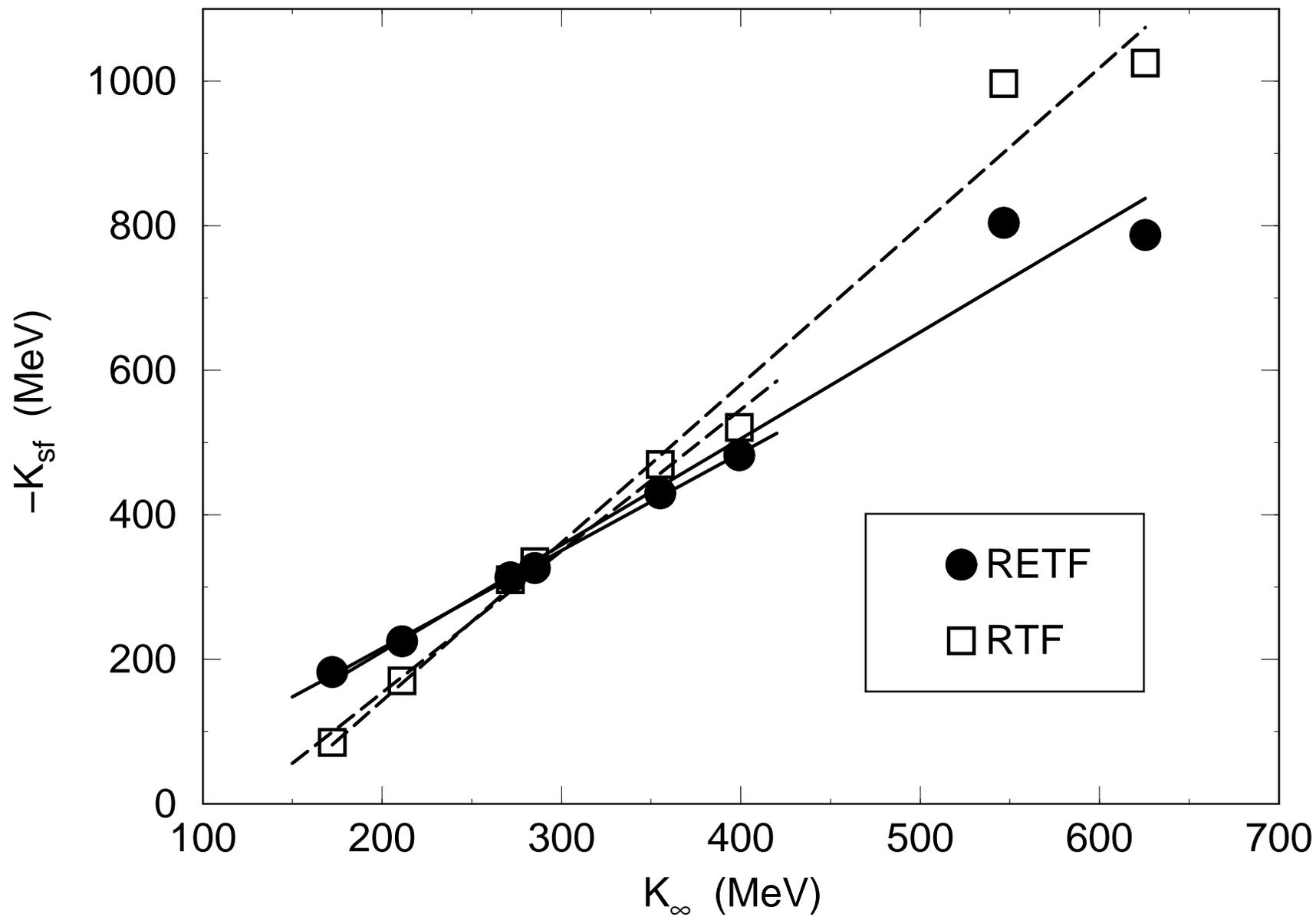
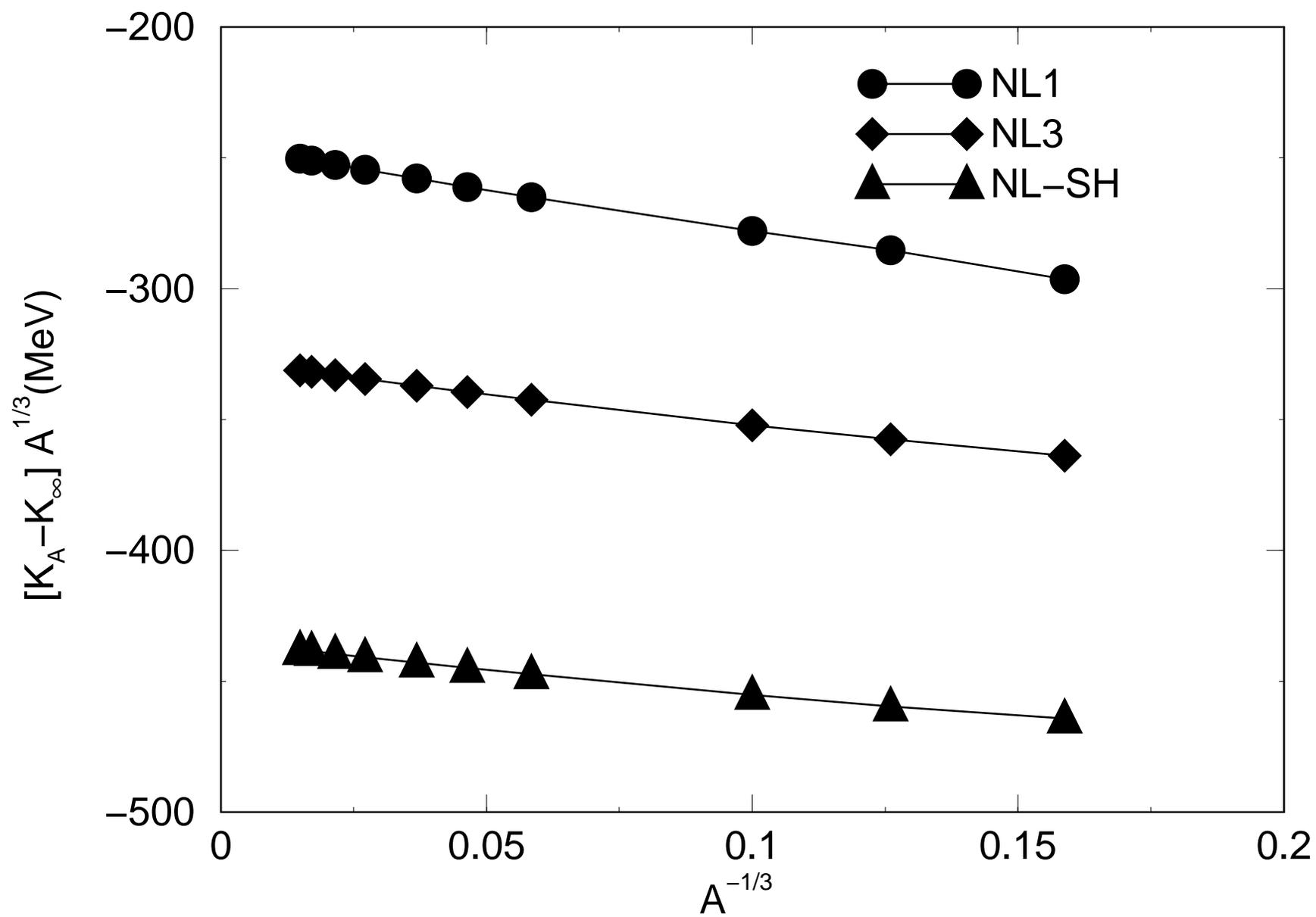
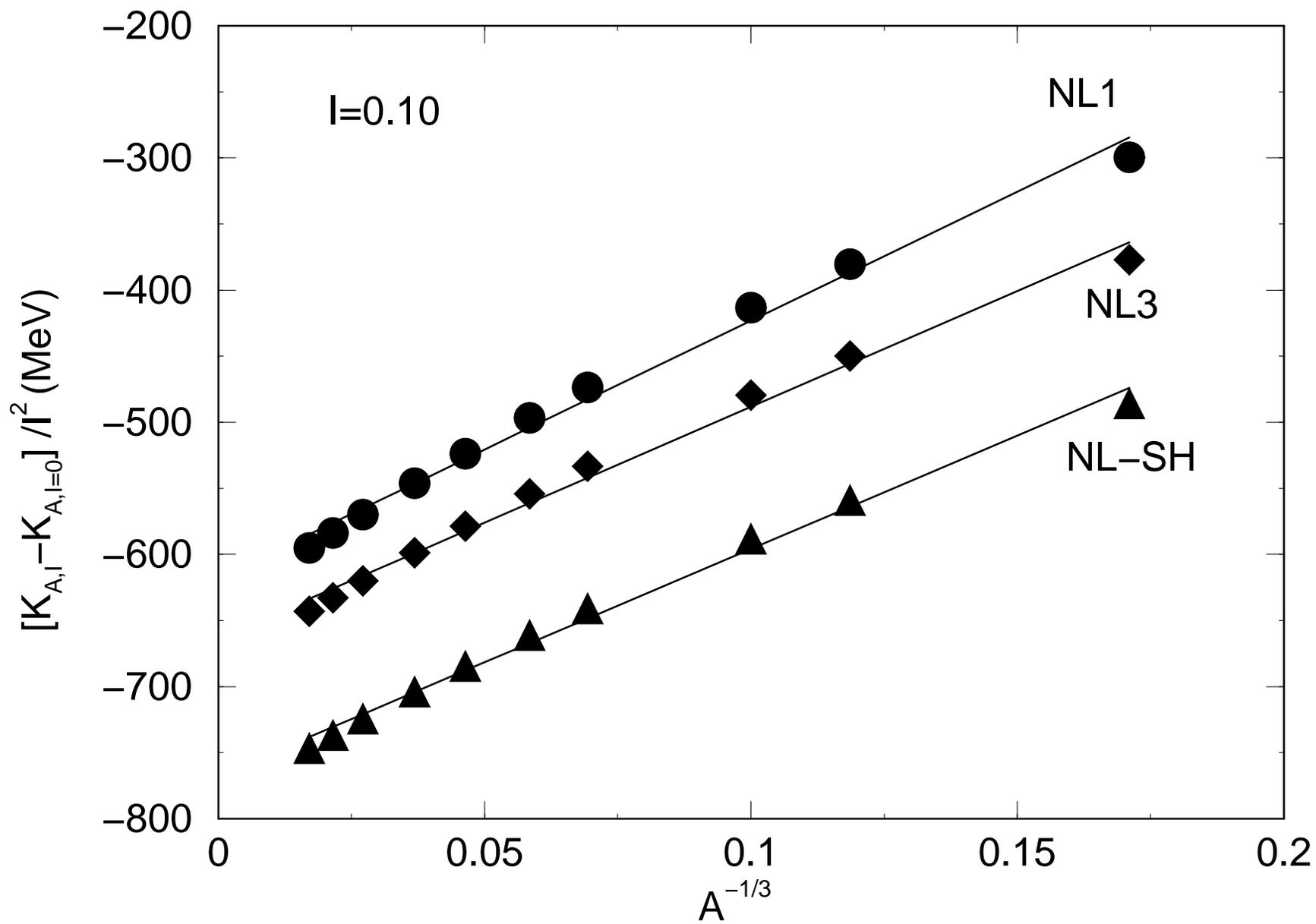
\end{document}